\documentclass[twocolumn,showpacs,preprintnumbers,amsmath,amssymb,footinbib]{revtex4}

\usepackage{graphicx}% Include figure files
\usepackage{epstopdf,mathrsfs}
\begin{document}

%\preprint{APS/123-QED}

\title{Spontaneous parametric processes in optical fibers: a comparison} % Force line breaks with \\

\author{Karina Garay-Palmett$^1$, Mar\'ia Corona$^{1,2}$, Alfred B. U'Ren$^1$\\
$^1$Instituto de Ciencias Nucleares, Universidad Nacional Aut\'onoma
de M\'exico\\ Apartado Postal 70-543, M\'exico 04510, DF\\$^2$Centro
de Investigaci\'on Cient\'{\i}fica
y de Educaci\'on Superior de Ensenada, \\
Apartado Postal 2732, Ensenada B.C., 22860, M\'exico\\e-mail:
karina.garay@nucleares.unam.mx, alfred.uren@nucleares.unam.mx}

\date{\today}% It is always \today, today,
% but any date may be explicitly specified

%\newcommand{\epsfg}[2]{\centerline{\scalebox{#2}{\epsfbox{#1}}}} %

\begin{abstract}
We study the processes of spontaneous four-wave mixing and of
third-order spontaneous parametric downconversion in optical fibers,
as the basis for the implementation of photon-pair and
photon-triplet sources. We present a comparative analysis of the two
processes including expressions for the respective quantum states
and plots of the joint spectral intensity, a discussion of
phasematching characteristics, and expressions for the conversion
efficiency. We have also included a  comparative study based on
numerical results for the conversion efficiency for the two sources,
as a function of several key experimental parameters.\\

Estudiamos los procesos de mezclado de cuatro ondas espont\'aneo y
de conversi\'on param\'etrica descendente de tercer orden
espont\'anea en fibras \'opticas, como base para la implementaci\'on
de fuentes de parejas y tripletes de fotones. Presentamos un
an\'alisis comparativo de los dos procesos, incluyendo expresiones
para los estados cu\'anticos respectivos y gr\'aficos de la
intensidad espectral conjunta, una discusi\'on de las
caracter\'isticas de empatamiento de fases, y expresiones para la
eficiencia de conversi\'on. Tambi\'en hemos inclu\'ido un estudio
comparativo, basado en resultados num\'ericos, de la eficiencia de
conversi\'on para los dos procesos, en funci\'on de diferentes
par\'ametros experimentales.

\end{abstract}


%\keywords{Photons nonclassical states, Quantum entanglement, Four-wave mixing}%Use showkeys class option if keyword
\maketitle

\section{Introduction}

Nonclassical light sources, and in particular photon-pair sources, have
become essential for testing the validity of
quantum mechanics~\cite{zeilinger99} and for the implementation of quantum-enhanced technologies such as
quantum cryptography, quantum computation and quantum
communications~\cite{kok07}. Photon pairs can be generated through spontaneous parametric processes,
in which classical electromagnetic fields illuminate optically non-linear
media.
Specifically, photon-pair sources are commonly based on the process of spontaneous
parametric down conversion (SPDC) in second order nonlinear
crystals~\cite{burnham70}. However, in the last decade there has
been a marked interest in the development of photon-pair sources based on
optical fibers~\cite{sharping01}. In fibers, the process
responsible for generating photon pairs is spontaneous four-wave
mixing (SFWM), which offers several significant advantages over SPDC, for example in terms of
the conversion efficiency~\cite{garay10}.  The third-order
non-linearity in optical fibers which makes SFWM possible can also
lead to
the generation of photon triplets through the process of
third-order spontaneous parametric down conversion (TOSPDC) \cite{corona11}.

Recently, we have studied spontaneous parametric
processes in optical fibers, including both SFWM photon-pair sources and TOSPDC photon-triplet sources. In the context of SFWM sources, we have
carried out a thorough theoretical study of the spectral correlation
properties between the signal and idler photons~\cite{garay07,garay08,garay08a}, which permits
tailoring these properties to the needs of specific quantum information
processing applications.   In particular, our results have helped pave the way towards the experimental realization of factorable photon-pair
sources~\cite{halder09,cohen09,soller10}, which represent an essential
resource for the implementation of linear optics quantum computation
(LOQC)~\cite{knill01}.  Likewise, we have analyzed the important
aspect of the attainable conversion efficiency, for the
pulsed-pumps and monochromatic-pumps regimes, as well as for
degenerate-pumps and non-degenerate-pumps configurations~\cite{garay10}.

Even though a number of approaches for the generation of photon
triplets have been either proposed or
demonstrated~\cite{persson04,keller98,rarity98,hubel10}, the
reported conversion efficiencies have been extremely low. Recently,
we have proposed a scheme for the generation of photon triplets in
thin optical fibers by means of TOSPDC~\cite{corona11}.  Our
proposed technique permits the direct generation of photon triplets,
without postselection, and results derived from our numerical
simulations have shown that the emitted flux for our proposed source
is competitive when compared to other proposals~\cite{hubel10}. Advances
in highly non-linear fiber technology are likely to enhance
the emission rates attainable through our proposed technique.

In this paper, we present a comparison of the SFWM
and TOSPDC processes. To this end, we assume a specific fiber
with a specific pump frequency which permits the realization of both processes.
We restrict our attention to SFWM involving degenerate pumps, and to TOSPDC
with degenerate emitted frequencies. Our comparison includes the
following aspects: i) the quantum state, leading to the joint
spectrum, ii) the phasematching properties, and iii) the conversion efficiency.

\section{Theory of spontaneous parametric processes in optical fibers}

Non-linear processes in optical fibers originate from the third
order susceptibility $\chi^{(3)}$ \cite{agrawal07}. Photon pairs
and triplets can be generated in optical fibers by means
of SFWM and TOSPDC, respectively.  Both of these processes, which
result from four wave mixing, require the fulfilment of energy and
momentum conservation between the participating fields. The current
analysis focuses on configurations in which all fields are linearly
polarized, parallel to the $x$-axis, and propagate in the same
direction along the fiber, which defines the $z$-axis.  Our work could
be generalized to cross-polarized source designs.

\begin{figure}[h!]
\begin{center}
\centering\includegraphics[width=8cm]{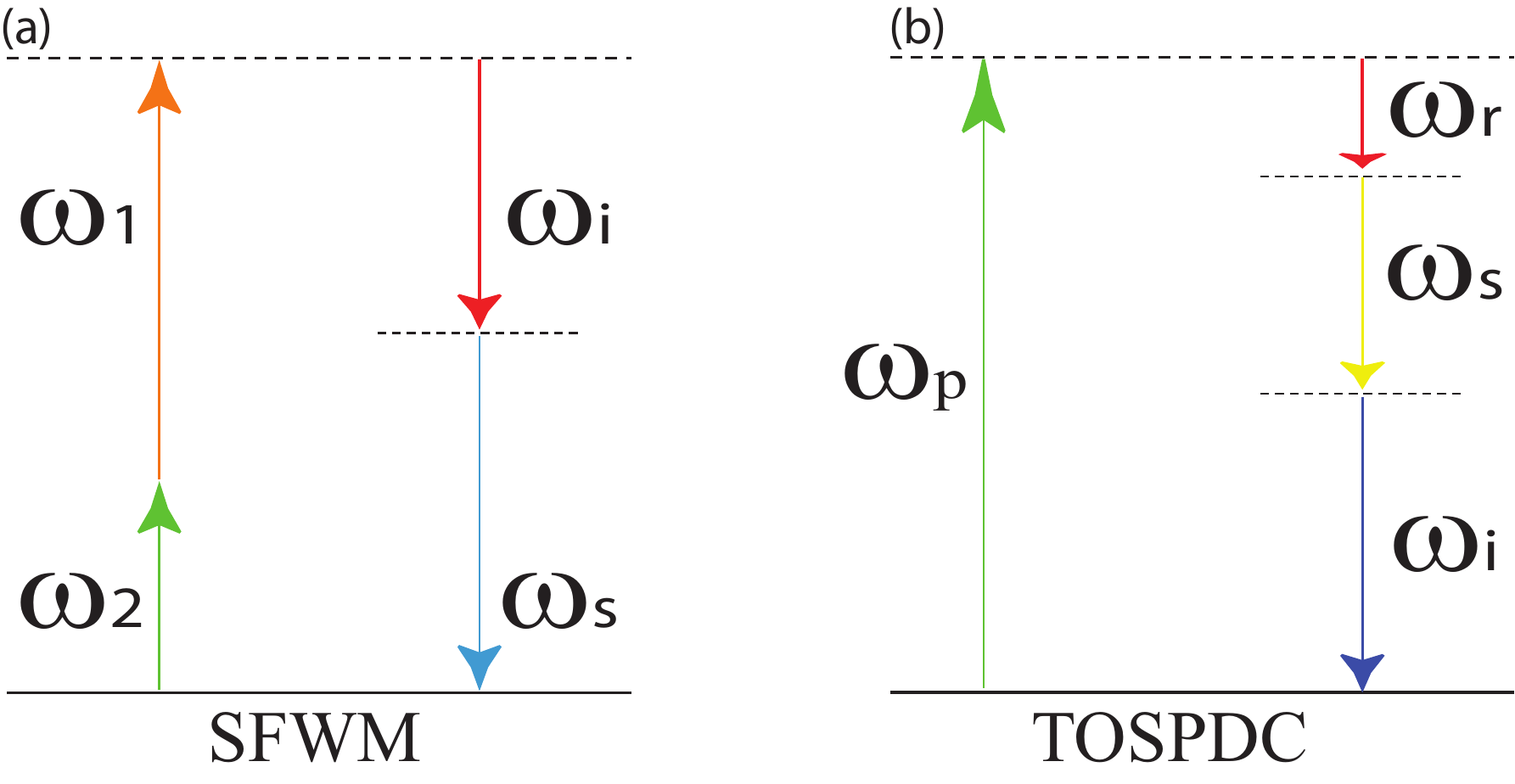}
\end{center}
\par
\caption{Energy level diagrams for the (a) SFWM and (b) TOSPDC
processes.} \label{levels}
\end{figure}

In the case of SFWM, two photons (one from each of two pump fields)
with frequencies $\omega_1$ and $\omega_2$  are jointly annihilated
giving rise to the emission of a photon pair, where the two photons
are typically referred to as signal(s) and idler(i), with
frequencies $\omega_s$ and $\omega_i$.  The energy conservation
relationship, thus reads $\omega_1+\omega_2=\omega_s+\omega_i$.  In
contrast, in the case of TOSPDC, a single pump photon at frequency
$\omega_p$ is annihilated, giving rise to a photon triplet, where
the three photons are here referred to as signal-1(r), signal-2(s)
and idler(i), with emission frequencies $\omega_r$, $\omega_s$ and
$\omega_i$.  The energy conservation constraint in this case reads
$\omega_p=\omega_r+\omega_s+\omega_i$. Representations of the SFWM
and TOSPDC processes, in terms of the frequencies involved, are
shown in Fig.~(\ref{levels}).

\subsection{Two-photon and three-photon quantum state}

Following a standard perturbative approach~\cite{mandel} we have
previously demonstrated that the SFWM
two-photon state~\cite{garay07} and the TOSPDC three-photon state~\cite{corona11,corona11b} can be written as

\begin{eqnarray}
\label{eq: state}   & &|\Psi\rangle = |0\rangle_{s}|0\rangle_{i}
+\kappa\!\!\int\!\! d\omega_{s} \!\!\int \!\! d\omega_{i}F\left(
\omega_{s},\omega_{i}\right) \left| \omega_{s}\right\rangle_{s}
\left|\omega_{i}\right\rangle_{i},
\end{eqnarray}

and

\begin{eqnarray}
\label{eq:state2}
|\Psi\rangle&=&|0\rangle_{r}|0\rangle_{s}|0\rangle_{i} +\zeta\int
\!\! d\omega_r  \!\!\int \!\! d\omega_{s}\nonumber
\\&\times&\int  \!\!  d\omega_{i}G\left(
\omega_r,\omega_{s},\omega_{i}\right) \left| \omega_{r}\right\rangle
_{r}\left| \omega_{s}\right\rangle _{s} \left|
\omega_{i}\right\rangle _{i},
\end{eqnarray}

\noindent respectively, where $\kappa$ and $\zeta$ are constants related to the
conversion efficiency.  In Eq.~(\ref{eq: state}), $F\left( \omega_{s},\omega_{i}\right)$ is
the SFWM joint spectral amplitude (JSA) function and is given by~\cite{garay10}

%In our analysis of SFWM and TOSPDC, we
%assume that the generated fields propagate in the same transverse
%fiber mode.   We assume that the pump field propagates
%in the same mode as the generated photons, for SFWM, or in a different mode
%compared to the generated modes, for TOSPDC, as dictated by phasematching needs.

\begin{eqnarray}
\label{JSA}F(\omega_s,\omega_i)&=&\int
d\omega\,\alpha_1(\omega)\alpha_2(\omega_s+\omega_i-\omega)\nonumber\\&\times&\mbox{sinc}\!\left[\frac{L}{2}\Delta
k(\omega,\omega_{s},\omega_{i})\right]e^{i\frac{L}{2}\Delta k(
\omega,\omega_{s},\omega_{i})},
\end{eqnarray}

\noindent where $L$ is the fiber length, $\alpha_{\nu}(\omega)$ is
the pump spectral envelope for mode $\nu=1,2$, and $\Delta k(\omega,\omega_{s},\omega_{i})$ is the
phasemismatch defined as

\begin{eqnarray}\label{delk}
\Delta k\left(  \omega,\omega_{s},\omega_{i}\right) &=&k_1\left(
\omega \right)  +k_2\left(  \omega_{s}+\omega_{i}-\omega\right)
-k\left(
\omega_{s}\right)  \nonumber\\& -&k\left(  \omega_{i}\right) -\Phi_{NL}, %
\end{eqnarray}

%\begin{eqnarray}\label{delk}
%\Delta k\left(  \omega,\omega_{s},\omega_{i}\right) &=&k_1\left(
%\omega \right)  +k_2\left(  \omega_{s}+\omega_{i}-\omega\right)
%-k\left(
%\omega_{s}\right)  \nonumber\\& -&k\left(  \omega_{i}\right) -\left(\gamma_1P_{1}+\gamma_2P_{2}\right), %
%\end{eqnarray}

\noindent which includes a nonlinear contribution
$\Phi_{NL}$ derived from self/cross-phase
modulation~\cite{garay07}.  It can be shown that

\begin{eqnarray}
\Phi_{NL}&=&(\gamma_1+2\gamma_{21}-2\gamma_{s1}-2\gamma_{i1})
P_1 \nonumber \\
&+&(\gamma_2+2\gamma_{12}-2\gamma_{s2}-2\gamma_{i2}) P_2
\end{eqnarray}

\noindent where
coefficients $\gamma_{1}$ and $\gamma_{2}$ result from self-phase
modulation of the two pumps, and are given with $\nu=1,2$ by

\begin{equation}\label{gam1}
\gamma_{\nu}=\frac{3\chi^{(3)}\omega_{\nu}^o}{4\epsilon_oc^2n_{\nu}^2A^{\nu}_{eff}}.
\end{equation}

In Eq.~\ref{gam1}, the refractive index $n_{\nu}\equiv n(\omega_{\nu}^o)$ and the effective
area $A_{eff}^{\nu}\equiv [\int\!\int\!dxdy|A_{\nu}(x,y)|^4]^{-1}$ (where the integral
is carried out over the transverse dimensions of the fiber)
are defined in terms of the carrier frequency $\omega_{\nu}^o$ for
pump-mode $\nu$~\cite{agrawal07}.  Here, functions
$A_{\mu}(x,y)$ (with $\mu=1,2,s,i$) represent the transverse field distributions
 and are assumed to be normalized such that
$\int\!\int\!dxdy|A_{\mu}(x,y)|^2=1$.

In contrast, coefficients $\gamma_{\mu\nu}$
($\nu=1,2$ and $\mu=1,2,s,i$) correspond to the cross-phase
modulation contributions that result from the dependence of the
refractive index experienced by each of the four participating fields on
the pump intensities.  These coefficients
are given by

\begin{equation}\label{gam2}
\gamma_{\mu\nu}=\frac{3\chi^{(3)}\omega_{\mu}^o}{4\epsilon_oc^2n_{\mu}n_{\nu}A^{\mu\nu}_{eff}},
\end{equation}

\noindent where $n_{\mu,\nu}\equiv n(\omega_{\mu,\nu}^o)$ is
defined in terms of the central frequency $\omega_{\mu,\nu}^o$ for
each of the four participating fields, and $A_{eff}^{\mu\nu}\equiv
[\int\!\int\!dxdy|A_{\mu}(x,y)|^2|A_{\nu}(x,y)|^2]^{-1}$ is the
two-mode effective overlap area  (note that
$A_{eff}^{\mu\nu}=A_{eff}^{\nu\mu}$).

Although in general terms $\gamma_{\nu}\neq\gamma_{\mu\nu}$ it may
be shown that for a SFWM interaction, the following represent valid
approximations: $\gamma_1 \approx \gamma_{21} \approx \gamma_{s1}
\approx \gamma_{i1}$ and $\gamma_2 \approx \gamma_{12} \approx
\gamma_{s2} \approx \gamma_{i2}$. Taking these approximations into account, we arrive
at the following simplified expression for $\Phi_{NL}$

\begin{equation}
\Phi_{NL}=\gamma_1P_{1}+\gamma_2P_{2}.
\end{equation}

The JSA function in Eq.~(\ref{JSA}) characterizes the spectral
entanglement present in the
SFWM photon pairs. We have previously
shown that depending on the type and degree of group velocity mismatch between the pump and the
emitted photons (which can be
controlled by tailoring the fiber dispersion), it becomes possible to generate two-photon states in a
wide range of spectral correlation regimes \cite{garay07}.

We now turn our attention to the three-photon TOSPDC state given by
Eq.~(\ref{eq:state2}), where $G(\omega_r,\omega_s,\omega_i)$
represents the TOSPDC three-photon joint spectral amplitude function.
This function characterizes the entanglement present in the photon
triplets and can be shown to be given by~\cite{corona11,corona11b}

\begin{equation}\label{Eq:FF}
G(\omega_r,\omega_s,\omega_i)=\alpha(\omega_r+\omega_s+\omega_i)\phi(\omega_r,\omega_s,\omega_i),
\end{equation}

\noindent where $\alpha(\omega_r+\omega_s+\omega_i)$ is the pump
spectral amplitude (PSA) and $\phi(\omega_r,\omega_s,\omega_i)$ is the
phasematching function (PM) which is given by

\begin{align} \label{fphmat}
\phi(\omega_r,\omega_s,\omega_i)&=\mbox{sinc}\left[L\Delta
k(\omega_r,\omega_s,\omega_i)/2\right]\nonumber\\&\times\exp[iL\Delta
k(\omega_r,\omega_s,\omega_i)/2],
\end{align}

\noindent written in terms of the fiber length $L$ and the
phasemismatch $\Delta k(\omega_r,\omega_s,\omega_i)$

\begin{align}\label{Eq:PM}
    \Delta k(\omega_r,\omega_s,\omega_i)&=
    k_p(\omega_r+\omega_s+\omega_i)-
    k_r(\omega_r)-k_s(\omega_s)\nonumber\\&-k_i(\omega_i)+[\gamma_p-2(\gamma_{rp}+\gamma_{sp}+\gamma_{ip})]P.
\end{align}

In Eq.~(\ref{Eq:PM}), the term in square brackets is the non-linear
contribution to the phase mismatch, where $P$ is the pump peak
power, and $\gamma_p$ and $\gamma_{\mu p}$ are the nonlinear
coefficients derived from self-phase and cross-phase modulation,
which are given by expressions of the same form as Eqns.~(\ref{gam1}) and (\ref{gam2}), respectively.

The joint spectral amplitude function for TOSPDC photon-triplets [see
Eq.~(\ref{Eq:FF})] is a clear generalization of the JSA which describes
photon-pairs generated by SPDC in second order nonlinear crystals~\cite{grice97}.
Note that while the TOSPDC JSA function is given as a simple product
of functions, the SFWM JSA function [see Eq.~(\ref{JSA})] is given by a
convolution-type integral, which has an exact solution for monochromatic pump
fields~\cite{garay08} and which likewise can
be integrated analytically for Gaussian-envelope pump fields, under
a linear approximation of
the phase mismatch of Eq.~(\ref{delk}) \cite{garay07}.

\subsection{Conversion efficiency in SFWM and TOSPDC processes}

A crucial aspect to consider in designing a photon-pair or
photon-triplet source is the conversion efficiency, to which we
devote this section. We present conversion efficiency expressions
previously derived by us, for the SFWM process~\cite{garay10} and
for the TOSPDC process~\cite{corona11,corona11b}, in terms of all
relevant experimental parameters.

Here, we define the conversion efficiency as the ratio
of the number of pairs or triplets emitted per unit time
to the number of pump photons per unit time.
In the case of pulsed pumps, we limit
our treatment to pump fields with a Gaussian spectral envelope,
which can be written in the form

\begin{equation}
\label{envSpec}\alpha_{\nu}(\omega)=\frac{2^{1/4}}{\pi^{1/4}\sqrt\sigma_{\nu}}\,\exp
\left[-\frac{(\omega-\omega_{\nu}^o)^2}{\sigma_{\nu}^2}\right],
\end{equation}

\noindent where $\omega_{\nu}^o$ represents the central frequency
and $\sigma_{\nu}$ defines the  bandwidth (with $\nu=1,2$).

We showed in Ref.~\cite{garay10} that the SFWM photon-pair
conversion efficiency can be written as

\begin{align}
\label{etapul}\eta&=\frac{2^8\hbar c^2n_1n_2}{(2\pi)^{3}R}
\frac{L^2\gamma_{fwm}^2p_1p_2}{\sigma_1\sigma_2(\omega_1^op_2+\omega_2^op_1)}\nonumber\\&
\times\int\!\!d\omega_s\!\!\int\!\!d\omega_i\,h_2(\omega_s,\omega_i)\left|f(\omega_s,\omega_i)\right|^2,
\end{align}

\noindent in terms of a version of the joint spectral amplitude [see
Eq.~(\ref{JSA})] defined as
$f(\omega_s,\omega_i)=(\pi\sigma_1\sigma_2/2)^{1/2}F(\omega_s,\omega_i)$,
which does not contain factors in front of the exponential and sinc
functions so that all pre-factors appear explicitly in
Eq.~(\ref{etapul}), and where the function $h_2(\omega_s,\omega_i)$
is given by

\begin{equation}
\label{hwswi}h_2(\omega_s,\omega_i)=\frac{k_s^{(1)}\omega_s} {
n_s^2}\frac{k_i^{(1)}\omega_i} { n_i^2},
\end{equation}

\noindent in terms of $k_{\mu}^{(1)}\equiv k^{(1)}(\omega_{\mu})$,
which represents the first frequency derivative of $k(\omega)$, and
where $n_{\mu}\equiv n(\omega_{\mu})$.

In Eq.~(\ref{etapul}), $\hbar$ is Planck's constant, $c$ is the
speed of light in vacuum, $p_{\nu}$ is the average pump power (for
$\nu=1,2$), $R$ is the pump repetition rate (we assume that two pump
fields have the same repetition rate), and the parameter
$\gamma_{fwm}$ is the nonlinear coefficient that results from the
interaction of the four participating fields and is different from
the parameters $\gamma_1$ and $\gamma_2$  of Eq.~(\ref{gam1}). This
parameter can be expressed as

\begin{equation}
\label{gamma}\gamma_{fwm}=
\frac{3\chi^{(3)}\sqrt{\omega_1^o\omega_2^o}}{4\epsilon_oc^2n_1n_2A_{eff}},
\end{equation}

\noindent where $A_{eff}$ is the effective interaction area among
the four fields given by

\begin{equation}
\label{aeffe}A_{eff}=\frac{1}{\int\!\!dx\!\!\int\!\!dyA_1(x,y)A_2(x,y)A_s^*(x,y)A_i^*(x,y)}.
\end{equation}

In the monochromatic-pumps limit, i.e. $\sigma_{1,2} \rightarrow 0$,
it can be shown that Eq.~(\ref{etapul}) is reduced to~\cite{garay10}

\begin{eqnarray}
\label{eficw} \eta_{cw}&=&\frac{2^5\hbar c^2
n_1n_2}{\pi}\frac{L^2\gamma_{fwm}^2p_1p_2}{p_1\omega_2+p_2\omega_1}\nonumber\\&\times&\!\!\!\!\int\!\!d\omega
h_2(\omega,\omega_1+\omega_2-\omega)\mbox{sinc}^2[L\Delta
k'_{cw}(\omega)/2].
\end{eqnarray}

\noindent where
$h_2(\omega,\omega_1+\omega_2-\omega)$ is given according to
Eq.~(\ref{hwswi}), and where the phase mismatch $\Delta
k'_{cw}(\omega)=\Delta k_{cw}(\omega,\omega_1+\omega_2-\omega)$ is
written in terms of the function

\begin{eqnarray}
\label{Dkcw} \Delta
k_{cw}(\omega_s,\omega_i)&=&k\left[\left(\omega_s+\omega_i+\omega_{1}
-\omega_{2}\right)/2\right]\nonumber\\&+&k\left[\left(\omega_s+\omega_i-\omega_{1}+
\omega_{2}\right)/2\right] \nonumber\\&-&k(\omega_s)-k(\omega_i)-(
\gamma_1p_1+\gamma_2p_2).
\end{eqnarray}

It is clear from Eqns.~(\ref{etapul}) and (\ref{eficw}) that the
SFWM conversion efficiency has a linear dependence on pump power,
or alternatively the emitted flux has a
quadratic dependence on this parameter. Note that although the
phasemismatch has a pump-power dependence, no
deviation from the linear behavior is
observed for power levels considered as typical. Note that these
conversion efficiency expressions are valid only in the spontaneous
limit, where the pump powers are low enough to avoid generation
events involving multiple pairs.

As is also clear form Eqns.~(\ref{etapul}) and (\ref{eficw}), the
conversion efficiency has a quadratic dependence on the nonlinearity
coefficient $\gamma_{fwm}$, which implies that it has an inverse
fourth power dependence on the transverse mode radius
\cite{agrawal07}. It can be shown that in general the double
integral in Eq.~(\ref{etapul}), or the single integral in
Eq.~(\ref{eficw}), scales as $L^{-1}$, so that taking into account
the $L^2$ appearing as a prefactor, the conversion efficiency scales
linearly with $L$. Likewise, it can be shown that in general the
double integral in Eqns.~(\ref{etapul}) scales as $\sigma^{3}$, so
that $\eta$ in Eq.~(\ref{etapul}) has a linear dependence on the
pump bandwidth.

In what follows, we focus on the conversion efficiency for the TOSPDC
process. As we have shown for the pulsed-pump regime
[see Eq.~(\ref{envSpec})], this efficiency
can be written as~\cite{corona11}

\begin{eqnarray}\label{efftri}
    \eta &=&\frac{2^{5/2}3^{2}c^{3}\hbar^{2}n_p^{3}}{(\pi)^{5/2}\omega_{p}^o}\frac{L^{2}\gamma_{pdc}^{2}}{\sigma}\nonumber\\&\times&\!\!\!\!\int\!\! d\omega_r\!\!  \int\!\!  d\omega_s\!\!  \int\!\!   d\omega_i
    \,h_3(\omega_r,\omega_s,\omega_i)|g(\omega_r,\omega_s,\omega_i)|^{2}.
\end{eqnarray}

\noindent which is given in terms of the function

\begin{equation} \label{Eq:hfunct}
    h_3(\omega_r,\omega_s,\omega_i)= \frac{k_r^{(1)}\omega_r}{n_r^2}\frac{k_s^{(1)}\omega_s}{n_s^2}
    \frac{k_i^{(1)}\omega_i}{n_i^2},
\end{equation}

\noindent and the new function
$g(\omega_r,\omega_s,\omega_i)=(\pi\sigma^{2}/2)^{1/4}G(\omega_r,\omega_s,\omega_i)$,
which is a version of the joint spectral amplitude
$G(\omega_r,\omega_s,\omega_i)$  [see Eq.~(\ref{Eq:FF})], which does
not contain factors in front of the exponential and sinc functions,
so that all pre-factors terms appear explicitly in
Eq.~(\ref{efftri}).

In Eq.~(\ref{efftri}) $\gamma_{pdc}$ is the nonlinear coefficient
that governs the TOSPDC process, given by

\begin{equation} \label{Eq:gam3}
    \gamma_{pdc}=\frac{3\chi^{(3)}\omega_{p}^o}{4\epsilon_{0}c^{2}n_p^{2}A_{eff}},
\end{equation}

\noindent with $A_{eff}$ the effective interaction area among the
four fields, which is expressed as $\left[\int dx\int dy A_p(x,y)
A_r^{*}(x,y)A_s^{*}(x,y)A_i^{*}(x,y)\right]^{-1}$, where the integral is carried out
over the transverse dimensions of the fiber.  Note that this
nonlinear coefficient is different from parameters $\gamma_{\nu}$
and $\gamma_{\mu\nu}$ defined in Eqs.~(\ref{gam1}) and (\ref{gam2}),
respectively.

For a monochromatic pump, the
conversion efficiency can be obtained by taking
the limit $\sigma \rightarrow 0$ [see Eq.~(\ref{efftri})], from which we obtain

\begin{eqnarray}\label{eficwtri}
    \eta_{cw} &=&\frac{2^{2}3^{2}\hbar^{2} c^{3}n_p^{3}\gamma_{pdc}^{2}L^{2}}{\pi^2\omega_{p}}\nonumber \\&\times&
   \int\!\! d\omega_r\!\!\int\! \!d\omega_s h_3(\omega_r,\omega_s,\omega_{p}-\omega_r-\omega_s)
   \nonumber \\&\times& \mbox{sinc}^{2}\!\!\left(\!\frac{L}{2}\Delta k_{cw}\!\right),
\end{eqnarray}

\noindent where $h_3(\omega_r,\omega_s,\omega_{p}-\omega_r-\omega_s)$
is given according to Eq.~(\ref{Eq:hfunct}), where $p$ is the average pump power, and where the phasemismatch $\Delta k_{cw}(k_r,k_s)$ [see Eq.~(\ref{Eq:PM})] is given by

\begin{eqnarray}
    \Delta
    k_{cw}(\omega_r,\omega_s)&=&k_p(\omega_p)-k(\omega_r)-k(\omega_s)-k(\omega_{p}-\omega_r-\omega_s)\nonumber\\&+&[\gamma_p-2(\gamma_{rp}+\gamma_{sp}+\gamma_{ip})]p.
\end{eqnarray}

As in the case of SFWM, for TOSPDC the conversion efficiency [see
Eqns.~(\ref{efftri}) and (\ref{eficwtri})] has a quadratic
dependence on the nonlinear coefficient $\gamma_{pdc}$, which
implies an inverse fourth power dependence on the transverse mode
radius. Thus, for both processes small core radii favor a large
emitted flux.

An important difference between the two processes relates to the dependence of
the conversion efficiency on the pump power and bandwidth.
While the TOSPDC conversion efficiency is independent of the pump
power (except for the pump-power dependence of the phasemismatch,
which can be neglected for typical pump-power levels), see Eqns.~(\ref{efftri}) and
(\ref{eficwtri}), the SFWM conversion
efficiency scales linearly with the pump power.  Note that in this respect the
behavior for TOSPDC is identical to that observed for SPDC in second-order nonlinear crystals.
Underlying this behavior is the fact that for SFWM
two pump photons are annihilated per generation event,
while for TOSPDC, and for SPDC, a single pump photon is annihilated per generation event.

Likewise, it can be shown that for source designs regarded as typical, the triple integral in Eq.~(\ref{efftri})
scales linearly with the pump bandwidth $\sigma$ so that the TOSPDC conversion efficiency is constant
with respect to $\sigma$ (within the phasematching bandwidth)~\cite{comment1}.  Again, note that this
behavior is identical to that observed for SPDC.  This is
to be contrasted with the linear dependence of the SFWM conversion efficiency
with $\sigma$.   This implies that unlike SFWM sources, for TOSPDC sources a pulsed-pump configuration does
not represent an advantage vs a monochromatic-pump configuration in terms of the attainable emitted flux.
In fact, the conversion efficiency
for \textit{spontaneous} four wave mixing scales with pump
power and bandwidth in the same manner as for a \textit{stimulated}
process, such as second harmonic generation, based on a second order nonlinearity.
This implies that (for sufficiently high pump powers) SFWM sources can be considerably brighter than
both TOSPDC and SPDC sources.   As a concrete illustration, in a remarkable
recent SPDC experiment~\cite{Krischek10}, despite extensive source
optimization the observed photon-pair flux, per pump power and per unit emission
bandwidth, is $\sim 500$ times lower
compared to a representative SFWM experiment~\cite{fulconis07}.

Finally, it can be shown that for source designs regarded as typical, the frequency integrals in Eqns.~(\ref{efftri}) and (\ref{eficwtri})
scale
as $L^{-1}$, so that the TOSPDC conversion efficiency exhibits a linear
dependence on $L$ as in the case of SFWM.~\cite{comment2}

\section{Phase matching properties for SFWM and TOSPDC}

In this section, we describe the techniques studied by us, designed to achieve
phasematching for the SFWM and TOSPDC processes in fused-silica fibers. In both cases,
phasematching properties are linked to the frequency-dependence of
the propagation constant $k(\omega)$ for each of the four participating
fields.

On the one hand for SFWM we assume that all four fields propagate
in the same transverse fiber mode, in particular in the HE$_{11}$
fiber mode. Our treatment could be generalized to the case where the
fields propagate in arbitrary transverse modes~\cite{agrawal07}. On
the other hand, for TOSPDC we adopt a multi-modal phasematching
strategy where the pump propagates in a different mode compared to
the generated photon triplets. Note that the frequency-degenerate
low-pump-power phasematching condition for TOSPDC can be written as
follows: $k_p(3 \omega)=3 k(\omega)$.  Because of the large spectral
separation between the pump and the generated photons, $k(3 \omega)$
is considerably larger than $3 k(\omega)$, for most common
optical materials, characterized by normal dispersion.   We propose
to exploit the use of two different transverse modes in a thin fiber
guided by air, i.e. with a fused silica core and where the cladding
is the air surrounding this core.  In particular, we will assume
that while the TOSPDC photons all propagate in the fundamental mode
of the fiber (HE$_{11}$), the pump mode propagates in the first
excited mode (HE$_{12}$) \cite{grubsky07}. We have shown that for
the generation of photon-triplets at a particular degenerate
frequency there is a single core radius for which the phase matching
condition is fulfilled \cite{corona11}.  This scheme can be easily
generalized to non-frequency-degenerate TOSPDC.

%For this configuration,
%phase matching can be easily achieved in different spectral ranges
%by using  pumps, in both, the normal dispersion \cite{rarity05}
%regime, and the anomalous dispersion regime \cite{fiorentino02}.

\begin{figure}[h!]
\begin{center}
\centering\includegraphics[width=8.5cm]{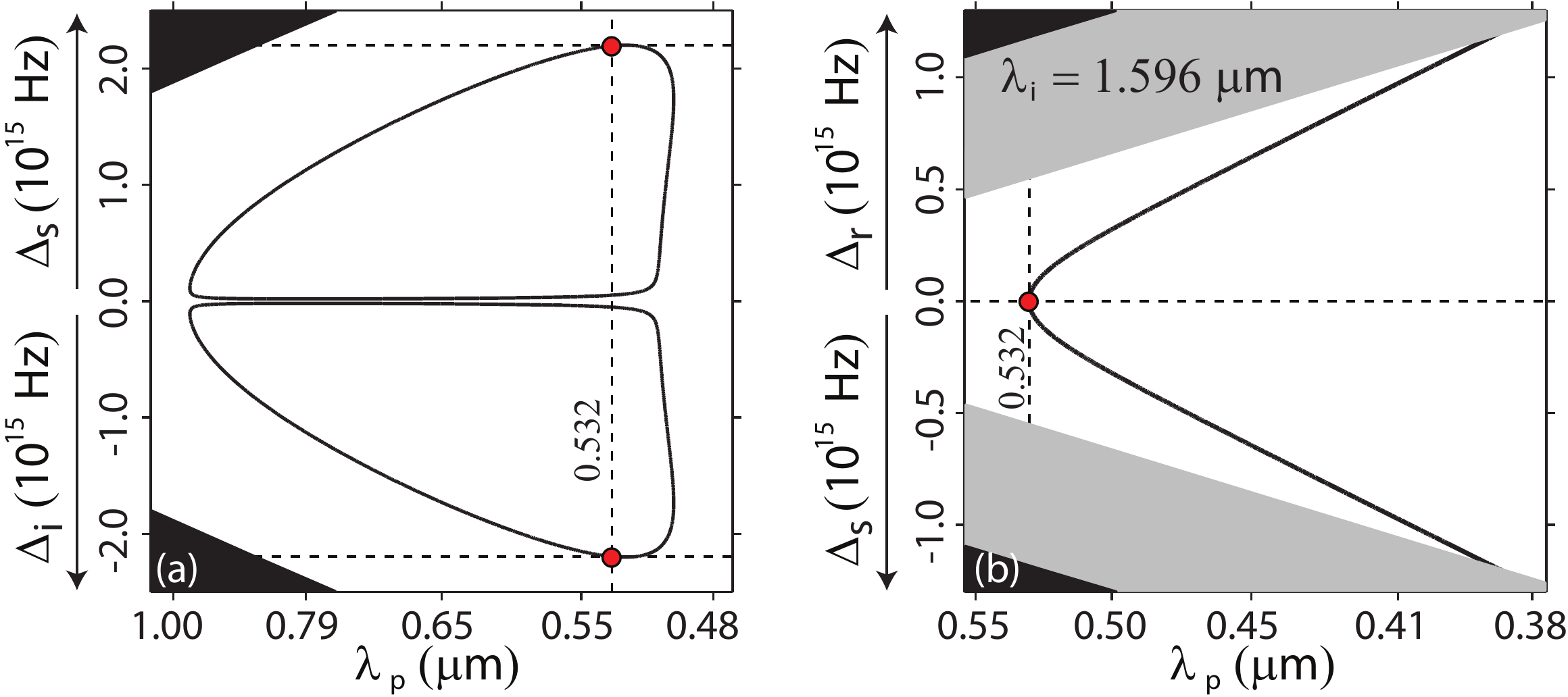}
\end{center}
\par
\caption{(a) Black, solid curve: perfect phasematching ($\Delta k=0$)
contour for degenerate-pumps SFWM . (b) Black, solid
curve: perfect phasematching ($\Delta k=0$) contour for TOSPDC with
$\lambda_i=1.596\mu$m. Black background: non-physical zone where
energy conservation would imply that one of the generated photons
has a negative frequency. Gray background: frequency zone outside of the
range of validity of the dispersion relation used for fused-silica.} \label{PMcont}
\end{figure}

In order to compare the two processes, we choose a single fiber
which can be used to implement both, a photon-pair SFWM source and a
photon-triplet TOSPDC source. We restrict this comparison to
degenerate-pumps SFWM and to TOSPDC involving frequency-degenerate triplets.
As a specific design, we consider a fiber guided by air with a core radius of $r=0.395\mu$m.
This core radius leads to TOSPDC phasematching involving a pump centered at $\lambda_p=0.532\mu$m and frequency-degenerate photon triplets centered at $\lambda=1.596\mu$m.
Fig.~\ref{PMcont} shows graphically the phasematching properties for the two
processes in terms of generation frequencies vs pump frequency (SFWM in panel A, and TOSPDC in
panel B). The black curves were  obtained by solving numerically, for each of
the two processes, the perfect phasematching condition.  We have displayed
the generation frequencies obtained assuming perfect phasematching in terms of detunings:
$\Delta_{s,i}=\omega_{s,i}-\omega_{p}$ for SFWM, and $\Delta_{r,s}=\omega_{r,s}-(\omega_{p}-\omega_{i})/2$
for TOSPDC ($\omega_\mu$, with $\mu=r,s,i,p$, represents the frequencies
for each of the participating modes). In the case of degenerate-pumps SFWM,
energy conservation dictates that $\Delta_{s}=-\Delta_{i}$ so that there are only
two independent frequency variables  ($\omega_p$ and  $\Delta_{s}$) and thus
Fig.~\ref{PMcont}(a) fully characterizes the relevant phasematching properties.
In the case of TOSPDC, in order to obtain a similar  representation of phasematching
properties we fix the idler-photon frequency [to $\omega_i=2\pi c/1.596\mu\mbox{m}$ in Fig.~\ref{PMcont}(b)], so that energy conservation dictates that $\Delta_{r}=-\Delta_{s}$. In this case, a series of plots similar to that in Fig.~\ref{PMcont}(b) each with a different value of $\omega_i$, is required for a full characterization of
the phasemathching properties.

In general, for a fiber exhibiting two zero-dispersion frequencies
within the spectral range of interest, the perfect phasematching contour in the space of generated
vs pump frequencies
is formed by two loops essentially contained between these two
zero dispersion frequencies~\cite{garay07}; this is illustrated in Fig.~\ref{PMcont}(a).
For a specific pump wavelength $\omega_p$ there may be two separate solutions for
$\Delta k=0$, leading to the inner and outer branches of the two loops.
However, the inner solutions tend to be spectrally near to
$\omega_p$, with $\Delta_s$ and $\Delta_i$ strongly determined by the nonlinear
contribution of the phasemismatch [see Eq.~(\ref{delk})], and thus pump-power dependent. This small
spectral separation can lead to contamination due to spontaneous
Raman scattering (which occurs within a window of $\sim 40$THz width towards
shorter frequencies from $\omega_p$).
In order to avoid Raman contamination, we exploit the outer branches
of the phasematching contour, which is comparatively less
dependent on the pump power. Note
that for this specific fiber, perfect phasematching occurs for pump
wavelengths within a range of approximately $470$nm. For
the photon-triplet source proposed in this study we have chosen a pump
wavelength of $\lambda_p=0.532\mu$m, that corresponds to the third
harmonic of $1.596\mu$m, which is the selected degenerate TOSPDC frequency. For this same fiber and for the same pump
wavelength, the SFWM process leads to signal and idler modes centered at
$0.329\mu$m and $1.398\mu$m, respectively. In fig.~\ref{PMcont}(a)
the selected pump wavelength and the corresponding signal and idler
frequencies are indicated by a black dashed line and red
circle dots, respectively.

Unlike for the SFWM process [see
fig.~\ref{PMcont}(a)], the perfect TOSPDC phasematching contour (with
$\omega_i$ kept constant) is an open curve where the vertex (red circle dot),
corresponds to frequency-degenerate
photon-triplet emission and where the selected pump frquency is indicated by a
vertical black-dashed line. It can be seen that
keeping $\omega_i$ constant at $\omega_p/3$, the pump can be tuned over a wide
frequency range, resulting in a wide tuning range for $\omega_r$ and $\omega_s$, away from $\omega_p/3$. It is worth mentioning that in general,
the nonlinear phasemismatch contribution
[see Eq.~(\ref{Eq:PM})] can be neglected for pump-power levels regarded
as typical.

\begin{figure}[h!]
\begin{center}
\centering\includegraphics[width=6cm]{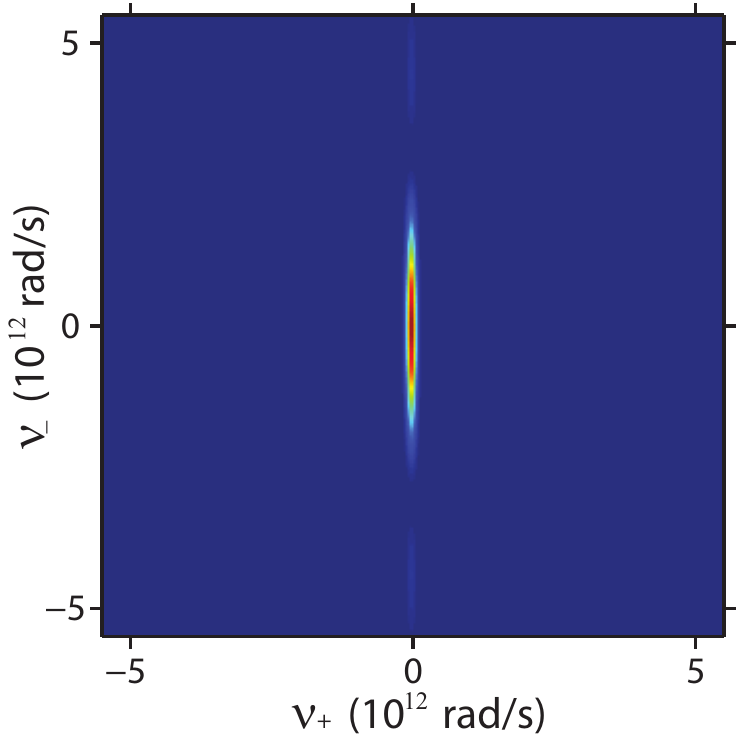}
\end{center}
\par
\caption{SFWM joint spectral intensity for SFWM photon pair state,
plotted as a function of frequency variables $\nu_+$ and $\nu_-$.} \label{JSI_pares}
\end{figure}

\begin{figure}[h!]
\begin{center}
\centering\includegraphics[width=8.5cm]{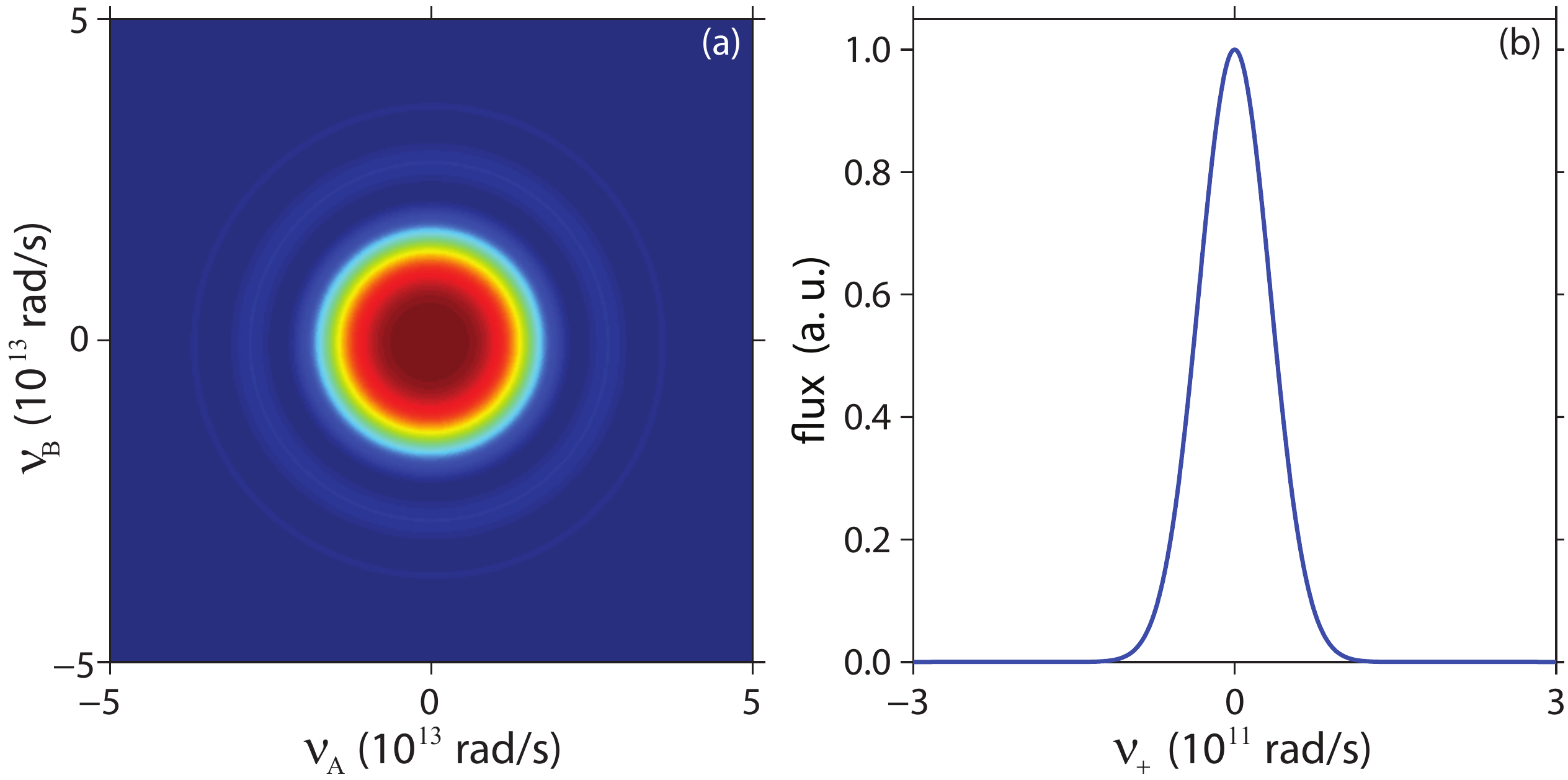}
\end{center}
\par
\caption{Representation of the TOSPDC joint spectral intensity
obtained for the same fiber and pump parameters as in
Fig.~\ref{JSI_pares}. (a) JSI evaluated at $\nu_{+}=0$. (b) JSI
evaluated at $\nu_{A}=\nu_{B}=0$.} \label{stateTOSPDC}
\end{figure}

In Figs.~\ref{JSI_pares} and \ref{stateTOSPDC} we show plots of the
joint spectral intensity (JSI) function, for the SFWM and TOSPDC
sources implemented with the specific fiber described above.  These
JSI functions are given by $|F(\omega_s,\omega_i)|^2$ and
$|G(\omega_r,\omega_s,\omega_i)|^2$, respectively. If properly
normalized, the JSI represents the probability distribution
associated with the different emission frequencies.

A plot of the JSI shows the type and degree of spectral correlations which
underlie the spectral entanglement present in the photon pairs or triplets.
Typical spectral correlations imply that, for both SFWM and TOSPC,
the JSI is tilted in the generated frequencies space, with narrow spectral features along specific directions.   Thus, for the fiber
parameters which we have assumed, the SFWM JSI exhibits a narrow width along the
$\omega_s+\omega_i$ direction, and a much larger width in the perpendicular direction.
In the case of TOSPDC, the JSI exhibits a narrow width along the
$\omega_{r}+\omega_{s}+\omega_{i}$ direction and much larger widths along the perpendicular
directions.  This means that, for both processes, it is convenient to plot the JSI in frequency variables chosen in accordance
with the correlations present.

Figure~\ref{JSI_pares} shows the JSI for the SFWM source, plotted vs
$\nu_{+}=\frac{1}{\sqrt{2}}(\nu_s+\nu_i)$ and
$\nu_{-}=\frac{1}{\sqrt{2}}(\nu_s-\nu_i)$, defined in terms of
frequency detunings $\nu_s\equiv \omega_s-\omega_{s}^0$ and
$\nu_i\equiv \omega_i-\omega_{i}^0$ where $\omega_{s}^0$ and
$\omega_{i}^0$ represent signal and idler frequencies for which
perfect phasematching is obtained.   For this plot, we have assumed
a fiber length of $L=1$cm and a pump bandwidth of $\sigma=0.118$THz
(which corresponds to a Fourier-transform-limited pulse duration of
$20$ps). The figure reveals that for this specific parameter
combination, the signal and idler photons are spectrally
anti-correlated.

Fig.~\ref{stateTOSPDC} shows a representation of
the three-photon TOSPDC JSI, where we have assumed the same
values for the fiber length and pump bandwidth that we used for SFWM,
plotted as a function of the following frequency
variables

\begin{align}
\nu_{+}&=\frac{1}{\sqrt{3}}(\omega_r+\omega_s+\omega_i-3\omega_0)  \nonumber \\
\nu_A&=\frac{1}{2}\left(1-\frac{1}{\sqrt{3}}\right) \omega_r + \frac{1}{2}\left(-1-\frac{1}{\sqrt{3}}\right) \omega_s+\frac{1}{\sqrt{3}} \omega_i \nonumber \\
\nu_B&=\frac{1}{2}\left(1+\frac{1}{\sqrt{3}}\right) \omega_r + \frac{1}{2}\left(-1+\frac{1}{\sqrt{3}}\right) \omega_s-\frac{1}{\sqrt{3}} \omega_i.
\end{align}

\noindent where $\omega_0$ is  defined as $\omega_0 \equiv
\omega_p/3$. Note that the variable $\nu_+$ defined for TOSDPC is
different to that defined for SFWM, in both cases given in terms of
the sum of the generated frequencies.  In Fig.~\ref{stateTOSPDC}(a),
we have plotted the JSI in these new variables, evaluated at
$\nu_{+}=0$, and in Fig.~\ref{stateTOSPDC}(b) we have plotted the
JSI in these new variables, evaluated at $\nu_{A}=\nu_{B}=0$.  Note
that the width along $\nu_{+}$ is much narrower compared to the
width along $\nu_{A}$ and $\nu_{B}$, an indication of the existence
of spectral correlations.   The ratio of the width along $\nu_{A}$
or $\nu_{B}$ to the width along $\nu_+$ is an indication of the
strength of the correlations.

\section{SFWM and TOSPDC conversion efficiency for specific source designs}

In this section, we present numerical simulations of the expected
conversion efficiency as a function of various experimental
parameters (fiber length, pump power and pump bandwidth) for the
specific SFWM and TOSPDC sources described in the previous section
(see Figs.~\ref{PMcont}, \ref{JSI_pares} and \ref{stateTOSPDC}). We
include in our analysis both, the pulsed- and monochromatic-pump
regimes.  In order to make this comparison as useful as possible,
both sources are based on the same fiber (guided by air with a core
radius of $r=0.395\mu$m) and the same pump frequency
($\lambda_p=0.532\mu$m). While in the SFWM source the signal and
idler modes are centered at non-degenerate frequencies
($\lambda_s=0.329\mu$m and $\lambda_i=1.398\mu$m), the TOSPDC source
is frequency degenerate at $\lambda=1.596\mu$m.

For the SFWM source, the nonlinear coefficient $\gamma_{fwm}$ was
numerically calculated from Eq.~(\ref{gamma}) yielding a value of
$\gamma_{fwm}=629\mbox{(kmW)}^{-1}$.  The corresponding value for
the TOSPDC source, numerically-calculated from Eq.~(\ref{Eq:gam3}),
yields a value of $\gamma_{pdc}=19\mbox{(kmW)}^{-1}$ . Although the
two processes take place in the same fiber with the same pump
frequency, the striking difference in the nonlinear coefficient
results from the far superior overlap between the four participating
fields in case of the SFWM source, for which the four fields
propagate in the same fiber mode (HE$_{11}$). Taking into account
the quadratic dependence of the conversion efficiency (observed for
both processes) on
the nonlinearity, this clearly favors a greater brightness for the
SFWM source compared to the TOSPDC source.

\subsection{Pump bandwidth dependence} \label{bandw}

We will first consider the conversion efficiency for the two sources
described above as a function of the
pump bandwidth (while maintaining the energy per pulse, or
alternatively, the average power and the repetition rate
constant). For this analysis, we assume a fiber length of $L=1$cm, a
repetition rate $R=100$MHz and an average pump power $p=180$mW for both sources.
Note that as $\sigma$ varies, the temporal duration
varies, and consequently the peak power varies too.

\begin{figure}[h!]
\begin{center}
\centering\includegraphics[width=8.7cm]{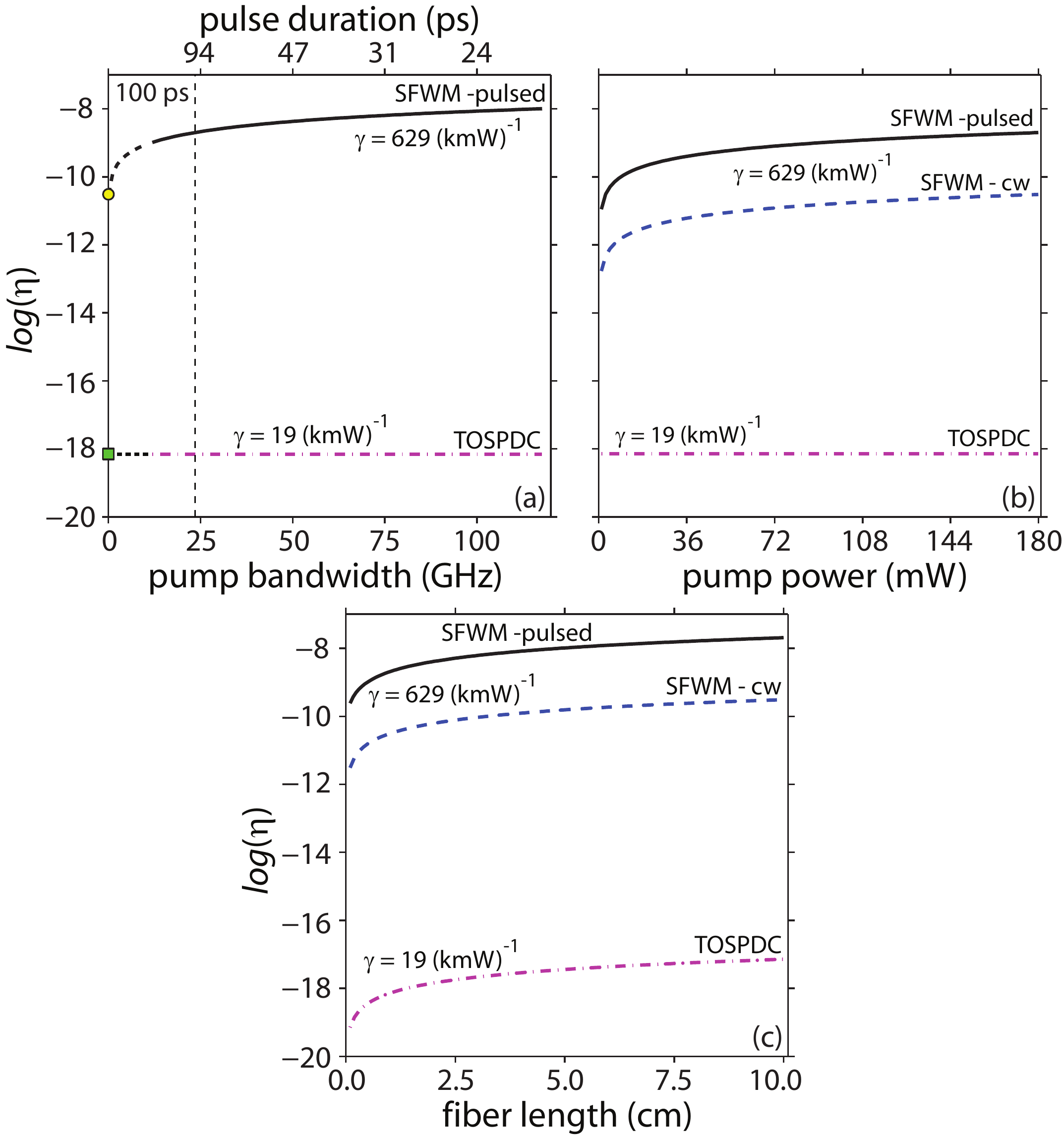}
\end{center}
\par
\caption{SFWM and TOSPDC conversion efficiency (in logarithmic scale) for the pulsed and
monochromatic pump regimes, as a function of: (a) the pump bandwidth
(the yellow circle-dot and the green square dot correspond to the
monochromatic-pump limit for SFWM and
TOSPDC, respectively), (b) the average pump power, and (c) the fiber
length.} \label{efficiency}
\end{figure}

We evaluate the conversion efficiency from Eqs.~(\ref{etapul}) and
(\ref{efftri}) for a pump bandwidth $\sigma$ range $23.5-117.7$GHz
(or a Fourier-transform-limited temporal duration
range $20-100$ps). Numerical results for the SFWM source [obtained
from Eq.~(\ref{etapul})] and for the TOSPDC source
[from Eq.~(\ref{efftri})] are shown in Fig.~\ref{efficiency}(a) (indicated by
the black solid line and the magenta dashed-dotted line,
respectively). The conversion efficiency has been plotted in a logarithmic scale,
considering the striking difference in order of magnitude between the efficiencies
for the two processes. It can be seen that
for the largest $\sigma$ considered, the SFWM conversion
efficiency is ten orders of magnitude greater than the TOSPDC conversion
efficiency. As expected, $\eta$ as given
by Eq.~(\ref{etapul}), exhibits a linear dependence on the pump
bandwidth (this is not graphically evident in the figure due to the logarithmic scale).  The black solid line in Fig.~\ref{efficiency}(a) shows
this behavior. Thus, for SFWM, the use of a pulsed pump significantly enhances the emitted flux
over the level attainable for the monochromatic-pump regime.
In contrast, the TOSPDC conversion efficiency
remains constant over the full range of pump
bandwidths considered. For this reason, in the case of TOSPDC, no difference
is expected in the emitted flux, between the monochromatic- and pulsed-pump
regimes (while maintaining the average pump power constant).

In the monochromatic-pump regime, evaluation of the SFWM conversion
efficiency through Eq.~(\ref{eficw}) predicts a
value of $\eta_{cw}=3.05\times10^{-11}$ [indicated by a yellow circle in
Fig.~\ref{efficiency}(a)]. Likewise, we calculate the TOSPDC conversion efficiency through
Eq.~(\ref{eficwtri}), from which we
obtain $\eta_{cw}=7.10\times10^{-19}$. This value is represented
in Fig.~\ref{efficiency}(a) by the green square. It is
graphically clear that the conversion efficiency values for $\sigma
\neq 0$ [calculated from Eq.~(\ref{etapul}) and Eq.~(\ref{efftri})]
approach the corresponding values in the monochromatic-pump limit [calculated from
Eq.~(\ref{eficw}) and Eq.~(\ref{eficwtri})].

\subsection{Pump power dependence}\label{pot}

We now turn our attention to the pump-power dependence of the conversion efficiency
for the two processes, while
maintaining the pump bandwidth and
other source parameters fixed. We compute the conversion efficiency as a function
of the average pump power, which is varied between $1$ and $180$mW.
We assume a fiber length of $L=1$cm,
a pump bandwidth of $\sigma=23.5$GHz (for the pulsed-pump case, corresponding to a
Fourier-transform-limited temporal duration of $100$ps) and a
repetition rate of $R=100$MHz.

Plots obtained numerically from our expressions [Eqs.~(\ref{etapul}) and
(\ref{efftri})] are presented in Fig.~\ref{efficiency}(b), where
$\eta$ is expressed in a logarithmic scale. The black solid line and
the magenta dashed-dotted line correspond to SFWM and TOSDPC,
respectively.  The SFWM conversion
efficiency in the monochromatic pump limit is obtained through
Eq.~(\ref{eficw}) and is indicated in
Fig.~\ref{efficiency}(b) by the blue dashed line.
As expected,
the SFWM conversion efficiency is considerably higher in
the pulsed-pump regime than in the monochromatic-pump regime.
Note that TOSPDC efficiency values, obtained from Eq.~(\ref{eficwtri}) for
the monochromatic-pump regime, are coincident with those obtained through Eq.~(\ref{efftri})
for the pulse-pump regime (see the discussion in the previous
subsection).

Fig.~\ref{efficiency}(b) shows that the SFWM conversion
efficiency is linear with pump power (which is not graphically evident due to the
logarithmic scale).  Note
that this linear dependence becomes quadratic for the flux vs average pump power.
For the
TOSPDC process, the situation is different: the conversion efficiency is constant with respect to the average pump power,
while the emitted flux varies linearly with the
pump power. As has already been remarked, this behavior is related to the fact that
two pump photons are annihilated per generation event for SFWM, while a single pump
photon is annihilated per generation event for TOSDPC.
In fact, this represents one of the essential advantages of SFWM
over SPDC photon-pair sources in terms of the possibility of obtaining
a large emitted flux, for sufficiently high pump powers.
Note that the process of TOSPDC
has important similarities with the process of SPDC;
in both cases, the
conversion efficiency is constant with respect to the pump power and to the
pump bandwidth (within the phasematching bandwidth).

At the highest average pump power considered, Eq.~(\ref{etapul})
predicts a SFWM conversion efficiency of $2.01\times10^{-9}$, which
can be contrasted with the value obtained in the monochromatic pump
limit through Eq.~(\ref{eficw}) ($\eta_{cw}=3.05\times10^{-11}$). In
turn, the TOSPDC conversion efficiency remains constant within the
full pump-power range considered with a value of
$7.11\times10^{-19}$, which is nine orders of magnitude lower than
the conversion efficiency of SFWM with a pulsed pump.

\subsection{Fiber length dependence}\label{Long}

We now turn our attention to the fiber-length dependence of the
conversion efficiency for the two processes, while maintaining other
source parameters fixed.  For this comparison we assume an average
pump power of  $p=180$mW and, for the pulsed case, a pump bandwidth
of $\sigma=23.5$GHz, and a repetition rate of $R=100$MHz. For this
study we vary the fiber length from $0.1$ to $10$cm, and as before
we assume a fiber radius of $r=0.395\mu$m;  recent experimental work
shows that it is possible to obtain a uniform-radius fiber taper of
$\sim 445$nm radius over a length of $9$cm\cite{lsaval04}. The
results obtained by numerical evaluation of Eqs.~(\ref{etapul}) and
(\ref{efftri}) in the pulsed-pump regime are shown graphically by
the black solid line for SFWM and by the magenta
dash-dot line  for TOSPDC.   The corresponding results obtained
for the monochromatic-pump regime by numerical evaluation of
Eqs.~(\ref{eficw}) and
(\ref{eficwtri}) are presented in
Fig.~\ref{efficiency}(c) by the blue dashed line for SFWM, while the curve for
TOSDPC overlaps the curve calculated for the pulsed case (magenta dash-dot line). As
expected, the conversion efficiency  exhibits a linear dependence on
the fiber length for both processes (which is not evident graphically due to the logarithmic scale). For the longest fiber
considered ($L=10$cm), the SFWM conversion efficiency  is
$2.04\times10^{-8}$ for the pulsed-pump regime and
$\eta_{cw}=3.09\times10^{-10}$ for the monochromatic-pump regime,
while the TOSPDC conversion efficiency is $7.13\times10^{-18}$ (for
both the pulsed- and monochromatic-pump regimes). Thus, for this
specific fiber, pulsed-pumped SFWM leads to two orders of magnitude
greater conversion efficiency than monochromatic-pumped SFWM, while
it leads to nine orders of magnitude greater conversion efficiency
than TOSDPC.

\section{Conclusions}

In this paper we have presented a comparative analysis of
two different types of source based on spontaneous non-linear processes in optical fibers:
photon-pair sources based on spontaneous four wave mixing (SFWM), and
photon-triplet sources based on spontaneous third-order parametric downconversion.
We have restricted our study to degenerate-pumps SFWM and to TOSPDC involving frequency-degenerate
photon triplets.  Likewise, we have assumed that all participating fields for each of the two types
of source are co-polarized.

We have presented expressions for the quantum state of SFWM
photon-pairs and TOSPDC photon-triplets, and we have discussed
differences in terms of phasematching properties for the two
processes.  We have presented expressions for the expected source
brightness for both processes, and for both: the pulsed-pump and
monochromatic-pump regimes.   Likewise, we have presented plots of
the joint spectral intensity for both processes, which elucidate the
type and degree of spectral correlations which underlie the
existence of spectral entanglement in each of the two cases.  We
have also presented the results of a comparative numerical analysis
of the attainable source brightness for each of the two sources, as
a function of key experimental parameters including pump bandwidth,
pump power, and fiber length.

From our study it is clear that SFWM sources can be much brighter than TOSPDC sources.  This is due
on the one hand to the far better degree of overlap between the four participating modes which can be attained for
SFWM, for which all fields propagate in the same fiber mode (HE$_{11}$), unlike TOSPDC for which our phasematching
strategy requires the use of two different fiber modes. On the other hand, for sufficiently high pump powers, this is due to the fact that for SFWM the
conversion efficiency scales linearly with pump power and bandwidth while for TOSPDC the conversion efficiency remains constant with respect to these two parameters.  Thus, unlike the case of TOSPDC, the use of short pump pulses can significantly enhance the SFWM conversion efficiency.  We expect that these results will be of use for the design of the next-generation of photon-pair and photon-triplet sources for quantum-information processing applications.

\section*{Acknowledgments}
This work was supported in part by CONACYT, Mexico,  by DGAPA, UNAM
and by FONCICYT project 94142.


\begin{thebibliography}{99}                                                                                               %

\bibitem {zeilinger99} A. Zeilinger,  Rev. Mod. Phys. \textbf{71}, 288–297 (1999).

\bibitem {kok07}  P. Kok,  W.J. Munro, K. Nemoto, T.C. Ralph, J.P. Dowling, and  G.J. Milburn,  Rev. Mod. Phys. \textbf{79}, 135–-174 (2007).

\bibitem {burnham70}  D. C. Burnham and D. L. Weinberg,  Phys. Rev. Lett. \textbf{25}, 84--87 (1970).

\bibitem {sharping01}  J. E. Sharping, M. Fiorentino, A. Coker, P. Kumar, and R. S. Windeler,  Opt. Lett. \textbf{26}, 1048--1050 (2001).   %4

\bibitem{garay10} K. Garay-Palmett, A. B. U'Ren, and R. Rangel-Rojo,  Phys. Rev. A \textbf{82}, 043809
(2010).

\bibitem{corona11} M. Corona, K. Garay-Palmett, and A. B.
U'Ren, Opt. Lett. \textbf{36}, 190--192 (2011).

\bibitem{garay07} K. Garay-Palmett, H. J. McGuinness, Offir Cohen, J.S. Lundeen, R. Rangel-Rojo, M. G. Raymer, C.J. McKinstrie, S. Radic, A. B.
U'Ren, and I.A. Walmsley,  Opt. Express \textbf{15}, 14870 (2007).


\bibitem {garay08}K. Garay-Palmett, A. B. U'Ren, R. Rangel-Rojo, R. Evans, and S. Camacho-Lopez,  Phys. Rev. A
\textbf{78}, 043827 (2008).

\bibitem{garay08a} K. Garay-Palmett, R. Rangel-Rojo, and A. B. U'Ren,  J. Mod. Opt. \textbf{55}, 3121-3131 (2008). %28

\bibitem {halder09} M. Halder, J. Fulconis, B. Cemlyn, A. Clark, C. Xiong, W. J. Wadsworth, and J. G. Rarity, Opt. Express \textbf{17}, 4670--4676 (2009).%29

\bibitem {cohen09} O. Cohen, J. S. Lundeen, B. J. Smith, G. Puentes, P. J. Mosley, and I. A. Walmsley,  Phys. Rev. Lett. \textbf{102}, 123603 (2009).

\bibitem{soller10}C. S\"{o}ller, B. Brecht, P.J. Mosley, L.Y. Zang, A. Podlipensky, N.Y. Joly, P.St.J. Russell, C. Silberhorn,  Phys. Rev. A \textbf{81}, 031801(R) (2010). %30

\bibitem{knill01} E. Knill, R. Laflamme and G.J. Milburn,  Nature \textbf{409}, 46 (2001).

\bibitem{persson04} J. Persson, T. Aichele, V. Zwiller, L. Samuelson, and O. Benson,  Phys. Rev. B \textbf{69}, 233314 (2004)

\bibitem{keller98}  T. E. Keller, M. H. Rubin, and Y. Shih, Phys. Rev. A \textbf{57}, 2076 (1998).

\bibitem{rarity98}  J. Rarity, and P. R. Tapster,  Phys. Rev. A \textbf{59}, R35
(1999).

\bibitem{hubel10}  H. H\"{u}bel, D. R. Hamel, A. Fedrizzi, S. Ramelov, K. J. Resch, and T. Jennewein,  Nature \textbf{466}, 601 (2010).

\bibitem {agrawal07}G. P. Agrawal, \textit{Nonlinear Fiber Optics, 4th Ed.} (Elsevier, 2007).

\bibitem {mandel}L. Mandel and E. Wolf, \textit{Optical Coherence and Quantum Optics}\/ (Cambridge University Press, 1995).


\bibitem{corona11b} M. Corona, K. Garay-Palmett,  and A. B.
U'Ren,  To be submitted.

\bibitem{grice97} W.P. Grice, and I. A. Walmsley, Phys. Rev. A \textbf{56}, 1627
(1997).

\bibitem{comment1} Note that situations exist where the TOSPDC flux vs $\sigma$ is not constant; e.g. a sufficiently long fiber implies that the resulting narrow phasematching function determines the joint amplitude function, leading to a flux which scales as $\sigma^{-1}$.


\bibitem{Krischek10} R. Krischek, W. Wieczorek, A. Ozawa, N. Kiesel, P. Michelberger, T. Udem, and H. Weinfurter, Nat. Photon. \textbf{4}, 170–-173, (2010).

\bibitem{fulconis07} J. Fulconis, O. Alibart, W. J. Wadsworth, and J. G. Rarity,  New J. Phys. \textbf{9} 276 (2007).
\bibitem{comment2} Note that situations exist where the TOSPDC flux vs $L$ is not linear; e.g. a sufficiently short fiber implies that the pump envelope function determines the joint amplitude function, leading to a flux which scales as $L^2$.

\bibitem{grubsky07} V. Grubsky, and J. Feinberg,  Opt. Commun. \textbf{274}, 447 (2007).

\bibitem{lsaval04} S. Leon-Saval, T. Birks, W. Wadsworth, P. St. J. Russell, and M. Mason, ` Opt. Express \textbf{12}, 2864 (2004).






\end{thebibliography}
\end{document}